\documentstyle[sprocl]{article}

\def\be{\begin{equation}}
\def\ee{\end{equation}}
\def\bea{\begin{eqnarray}}
\def\eea{\end{eqnarray}}

\begin{document}

\title{GREEN'S FUNCTION MONTE CARLO APPROACH TO SU(3) YANG-MILLS THEORY
IN (3+1)D}  

\author{C. J. HAMER, M. SAMARAS AND R.J. BURSILL}

\address{
School of Physics,
University of New South Wales,\\
Sydney, 2006, Australia\\
E-mail: C.Hamer@unsw.edu.au
}


\maketitle\abstracts{ 
	A `forward walking' Green's Function Monte Carlo algorithm is
used to obtain expectation values for SU(3) lattice Yang-Mills theory in
(3+1) dimensions. The ground state energy and Wilson loops are
calculated, and the finite-size scaling behaviour is explored. Crude
estimates of the string tension are derived, which agree with previous
results at intermediate couplings; but more accurate results for larger
loops will be required to establish scaling behaviour at weak coupling.
}

\section{Introduction}
\label{sec:intro}

	Classical Monte Carlo simulations provide a very powerful and
accurate method for the study of Euclidean lattice gauge theories. In
the Hamiltonian formulation, on the other hand, the corresponding
quantum Monte Carlo methods have been somewhat neglected. Here we
present a study of SU(3) Yang-Mills theory in (3+1) dimensions, using
the Green's Function Monte Carlo approach.\cite{kalos}

	Heys and Stump \cite{heys1} and Chin et al. \cite{chin1} pioneered the 
use of ``Green's Function Monte Carlo'' (GFMC) or ``Diffusion Monte
Carlo'' techniques in Hamiltonian LGT, in conjunction with a
weak-coupling representation involving continuous gauge field link
variables. This was successfully adapted to non-Abelian Yang-Mills
theories,\cite{chin2,chin3} with no minus sign problem arising. In this
representation, however, one is simulating the wave function in gauge
field configuration space by a discrete ensemble or density of random
walkers: it is not possible to determine the derivatives of the gauge
fields for each configuration, or to enforce Gauss's law explicitly.
the ensemble always relaxes back to the ground state sector. 
In order to compute the string tension or mass gap,
one must measure an 
appropriate correlation function, and estimate the mass gap as the 
inverse of the correlation length. 
We have introduced the `forward-walking' technique, well-known in
many-body theory,\cite{kalos,kalos1} to measure the expectation values and
correlation functions. The technique has been demonstrated for the cases
of the transverse Ising model in (1+1)D,\cite{samaras} and the U(1) LGT in
(2+1)D.\cite{hamer3}

	Here we apply the technique for the first time to a non-Abelian
model, namely SU(3) Yang-Mills theory in (3+1)D. The ground state energy
and Wilson loop values are calculated, and approximate values are
extracted for the string tension in the weak-coupling regime.
Comparisons are made with earlier calculations, where they are
available.\cite{hamer4}

\section{Method}
\label{sec2}

\subsection{Lattice Hamiltonian}

	The Green's Function Monte Carlo formalism has been adapted to
SU(2) Yang-Mills theory by Chin, van Roosmalen, Umland and
Koonin,\cite{chin2} and sketched for the SU(3) case by Chin, Long and
Robson.\cite{chin3} 

The SU(3) lattice Hamiltonian is given by \cite{chin3} 
\begin{equation}
H = \frac{g^{2}}{2a}\{ \sum_{l} E^{a}_{l}E^{a}_{l} -
\frac{\lambda}{3}\sum_{p}Tr(U_{p} + U_{p}^{\dag})\} \label{1}
\end{equation}
where $E_{l}^{a}$ is a component of the electric field at link l,
$\lambda = 6/g^{4}$, the index $a$ runs over the 8 generators of SU(3),
and $U_{p}$ denotes the product of four link operators around an
elementary plaquette.
 We will
work with the dimensionless operator
\begin{equation}
H = \frac{1}{2}\sum_{l}E_{l}^{a}E_{l}^{a} -\frac{\lambda}{6} \sum_{p}
Tr(U_{p} + U_{p}^{\dag}) \label{3}
\end{equation}

	The link variables are elements of the group SU(3) in the
fundamental representation
\begin{equation}
U = \exp(-i\frac{1}{2}\lambda^{a}A^{a})
\end{equation}

\subsection{Green's Function Monte Carlo method}

	The Green's Function Monte Carlo method employs the operator
$\exp(-\tau(H-E))$, i.e. the time evolution operator in imaginary time,
as a {\it projector} onto the ground state $|\psi_{0} \rangle$:
\begin{equation}
 |\psi_{0} \rangle  \propto  
  \lim_{\Delta\tau \rightarrow 0, N\Delta\tau \rightarrow \infty}
e^{-N\Delta\tau(H-E)}|\Phi\rangle
\label{4}
\end{equation}
where $|\Phi \rangle$ is any suitable trial state. To procure some
variational guidance, one performs a ``similarity transformation" with
the trial wave function $\Phi$, and evolves the {\it product}
$\Phi|\psi_{0}\rangle$ in imaginary time. The heart of the procedure is
the calculation of the matrix element corrersponding to a single small
time step $\Delta\tau$. Chin et al \cite{chin2} show that 
\begin{eqnarray}
\langle {\bf x}'| \Phi e^{-\Delta\tau(H-E)}\Phi^{-1}|{\bf x}\rangle & =
& \prod_{l}\langle
U_{l}'|N\{ \exp (-\frac {1}{2} \Delta \tau E_{l}^{a}E_{l}^{a}) 
\exp[\Delta\tau E^{a}_{l}
(E^{a}_{l}\ln\Phi)]\}|U_{l}\rangle \nonumber\\
 & & \exp\{\Delta\tau [E-\Phi^{-1}H\Phi({\bf
x})]\} + O(\Delta\tau^{2}) \nonumber\\
 & \equiv & p({\bf x',x})w({\bf x}) + O(\Delta\tau^{2}) \label{eq5}
\end{eqnarray}
where ${\bf x} = \{U_{l}\}$ denotes an entire lattice configuration of
link fields.

	The product $\Phi|\psi\rangle$ is simulated by the density of an
ensemble of random walkers. At the kth. step, the
`weight' of each walker at ${\bf x}_{k}$ is multiplied by $w({\bf
x}_{k})$.
 The effect of $p({\bf x_{k+1},x_{k}})$ is to alter
each link variable $U_{l}$ in $\{{\bf x_{k}}\}$ to $U_{l}'$ by a
Gaussian random walk plus a ``drift step" guided by the trial wave
function:
\begin{equation}
U' = \Delta U U_{d} U
\end{equation}
where $U_{d} = \exp [i\frac{1}{2}\lambda^{a}(i \Delta \tau E^{a} \ln \Phi )]$
is the drift step, and $\Delta U$ is an SU(3) group element randomly chosen
from a
Gaussian distribution around the identity, with variance
$\langle\Delta s^{2}\rangle = 8\Delta\tau$ (i.e. $\Delta \tau$ for each
index {\it a}), where
\begin{equation}
\langle \Delta s^{2}\rangle \approx \sum_{a}A^{a}A^{a} = 8\Delta \tau,
\end{equation}
for small $A^{a}$.

	The simulation is carried out for a large number of iterations
$\Delta\tau$, until an equilibrium distribution $\Phi|\psi_{0}\rangle$ is
reached. The energy E in (\ref{eq5}) is adjusted after each iteration so as to
maintain the total ensemble weight constant. The average value of E can
then be taken as an estimate of $E_{0}$, the ground-state energy.

	As time evolves, the weights of some walkers grow larger, while
others grow smaller, which would produce an increased statistical error.
To avoid this, a ``branching" process is employed, whereby a walker with
weight larger than some threshold is split into two independent walkers,
while others with weights lower than another threshold are amalgamated.

\subsection{Trial Wave Function}

	The trial wave function is chosen to be the one-parameter
form \cite{chin3}
\begin{equation}
\Phi = \exp[\alpha\sum_{p}Tr(U_{p} + U_{p}^{\dag})]
\end{equation}

	Then the drift step for each linkis \cite{chin2}
\begin{equation}
U_{d} 
 =  \exp[-i \frac{\lambda^{a}}{2}A^{a}_{l}],
\end{equation}
\begin{equation}
A^{a}_{l} = -i \Delta \tau \frac{\alpha}{2}\sum_{p \in l}
Tr[\lambda^{a}U_{l} .. U_{4}^{\dag} - h.c.]
\end{equation}

	Finally, the trial energy factor is
\begin{eqnarray}
\Phi^{-1}H\Phi & = & \sum_{l}\{\frac{\alpha^{2}}{8}(\sum_{p \in l}
Tr[\lambda^{a}U_{l} .. U_{4}^{\dag} - h.c.])^{2} \\
 & & +(\frac{2\alpha}{3} - \frac{\lambda}{24})\sum_{p \in l} Tr(U_{p} +
U_{p}^{\dag})\} .
\end{eqnarray}

\subsection{Forward Walking estimates}
\label{subsec: for}

	The ``forward walking" technique is used to estimate expectation
values.\cite{kalos} Its application to the U(1) lattice gauge theory in
(2+1)D was discussed by Hamer et al.\cite{hamer3}
It is implemented for an operator Q (assumed diagonal, for simplicity) by \cite{kalos1}
recording the value $Q({\bf x}_{i})$ for each ``ancestor" walker at the
 beginning of a measurement;
propagating the ensemble as normal for $J$ iterations, keeping a record
 of the ``ancestor" of each walker in the current population;
and taking the weighted average of the $Q({\bf x}_{i})$ with respect to the weights
 of the descendants of ${\bf x}_{i}$ after the $J$ iterations, using 
sufficient iterations $J$ that the estimate reaches a `plateau'.

\section{Results}

	Simulations were carried out for LxLxL lattices up to L=8 sites,
using runs of typically 4000 iterations and an ensemble size of 250
to 1000 depending (inversely) on lattice size. 
 Time
steps $\Delta\tau$ of 0.01 and 0.05 ``seconds" were used, with each iteration
consisting of 5 sweeps and 1 sweep through the lattice, respectively,
followed by a branching process.
The first 400 iterations were discarded to allow for equilibration.
The data were block averaged over blocks of up to 256 iterations, to
minimize the effect of correlations on the error estimates.

	The results taken at $\Delta\tau = 0.01$ and $ \Delta\tau =
0.05$ were extrapolated linearly to $\Delta \tau = 0$. 
	The variational parameter $c$ was given values as
 used by Chin et al,\cite{chin3}
obtained from a variational Monte Carlo calculation. We checked that
these were approximately the optimum values for small lattices.

	Forward-walking measurements were taken over $J$ iterations, where $J$ ranged from
20 to 100, depending on the coupling $\lambda$. Ten separate
measurements were taken over this time interval, in order to check
whether the value measured by forward-walking had reached equilibrium. A
new measurement was started soon after the previous one had finished.

\subsection{Ground-state Energy}

The dependence of the ground-state energy per site on lattice size is 
illustrated in Figure 1, at two fixed
couplings $\lambda = 3.0$ and $\lambda = 5.0$. In the
``strong-coupling" case, $\lambda = 3.0$, it can be seen that the
results converge exponentially fast in $L$, whereas in
the ``weak-coupling" regime, $\lambda = 5.0$, the convergence is more
like $1/L^{4}$ at these lattice sizes. This behaviour merits some
further explanation. 

	A similar phenomenon occurs in the case of the U(1) theory in (2+1)D.\cite{hamer5,hamer3} In the
strong-coupling regime, where the mass gap is large, the usual exponential convergence occurs. In
the weak-coupling regime, however, where the mass gap M is very small, the finite-size scaling
behaviour for small lattice sizes is that of a massless theory, and it is only at much larger
lattice sizes $ L \approx 1/M$ that a crossover to exponential convergence occurs. 
An ``effective Lagrangian"
corresponding to free, massless gluons (non-interacting QCD) should describe the finite-size
behaviour in the present case, in line with the idea of asymptotic freedom. By analogy with the
(2+1)D case, we expect a $1/L^{4}$ dependence for the corrections to the ground-state energy per
site. We hope to pursue this analysis further at a later date.

	An anomalous feature in Figure 3b)  is that the $L = 8$ point lies well out of line
with the others. This occurs at other couplings also.
We suspect that the results for $L = 8$ are not reliable, and that the trial wave function will
have to be further improved to give reliable results for such large lattices. 

We have made estimates of the bulk limit, extrapolating mainly from the
smaller L values where possible.
The estimates for the bulk ground-state energy per site are graphed
as a function of coupling in Figure 2, where they are compared with previous
estimates \cite{hamer4} obtained by an `Exact Linked Cluster Expansion'
(ELCE) procedure, and with the
asymptotic weak-coupling series.\cite{hofsass} 
The Monte Carlo results agree very well with the ELCE estimates, and
appear to match nicely onto the expected weak-coupling behaviour for
$\lambda \geq 6$.

\subsection{Wilson Loops}

The forward-walking method was used to estimate values for the m x n
Wilson loops, $W(m,n)$. 
A graph of the `mean plaquette' $W(1,1)$ versus the
variational parameter $c$ is shown in Figure 3. A problem is immediately
apparent. The estimate for $W(1,1)$ is not independent of $c$, in fact it
depends linearly on $c$ over this range, and the size of the variation is
such that the probable systematic error due to the choice of $c$ is an
order of magnitude larger than the random statistical error in the
results. Thus it would be advantageous in future studies to put more
effort into improving the trial wave function, rather than merely
improving the statistics.

The finite-size behaviour for the Wilson loops is similar to that of the
ground-state energy. 
 The estimates for the mean plaquette in the bulk limit are graphed as a function of
coupling $\lambda$ in Figure 4, and compared with series estimates at
strong and weak coupling.\cite{hamer4,hofsass} The agreement is quite
good.

\subsection{String Tension}

	Having obtained estimates for the Wilson loop values on the bulk
lattice, one can extract estimates for the `spacelike' string tension
using the Creutz ratios:
\begin{equation}
Ka^{2} \simeq R_{n} = - \ln \left[\frac{W(n,n)W(n-1,n-1)}{W(n,n-1)^{2}}
\right]
\end{equation}

	The results are shown in Figure 5. Also shown in Figure 5 are
some previous estimates derived from the `axial' string tension,
obtained \cite{hamer4} 
  using an
`Exact Linked Cluster Expansion' (ELCE) method. 
The axial string tension $aT$ is calculated as an energy per link, and
must be converted to a dimensionless, `spacelike' tension by dividing by
the `speed of light' \cite{hasenfratz} c.
We have also used the weak-coupling relationship between the scales of
Euclidean and Hamiltonian lattice Yang-Mills theory calculated by
Hasenfratz et al \cite{hasenfratz} to plot the results against the Euclidean
coupling $\beta = 6/g^{2}_{E}$.

	It can be seen that the present GFMC results are in rough
agreement with the axial string tension results in the region $4 \leq
\beta \leq 5$, which is also the region where the `roughening' transition
occurs in the string tension.\cite{hamer4} For $\beta > 5$, however, the
Creutz ratio $R_{2}$ runs above the ELCE estimate, and shows no sign of
the expected crossover to an exponentially decreasing scaling behaviour
at $\beta \simeq 6$. We presume that this is a finite-size effect, and that
the Creutz ratios $R_{n}$ for larger $n$ will show a substantial decrease
in the `weak-coupling' regime $\beta \geq 6$. That is certainly the
pattern seen in the Euclidean calculations, or in the
U$(1)_{2+1}$ model.\cite{hamer3} Unfortunately, however, our present results
for the larger Wilson loops are not of sufficient accuracy to allow
worthwhile estimates of $R_{n}$ for $n \geq 2$.

\section{Conclusions}

	Some significant problems with the GFMC method have emerged from
this study. The `forward-walking' technique was introduced specifically
to avoid any variational bias from the trial wave function.\cite{kalos,kalos1} As it
turns out, however, the results for the Wilson loops show a substantial
dependence on the trial wave function parameter $c$.
 The systematic error due to this
dependence is an order of magnitude larger than the statistical error,
so it would pay to put more effort in future studies into improving the
trial wave function, rather than simply increasing the statistics.
Furthermore, the effective ensemble size decreases during each
measurement as the descendants of each `ancestor' state die out, and
this produces a substantial loss in statistical accuracy at weak
coupling, as well.

	It would be preferable if one were able to do away entirely with
all the paraphernalia of trial wave function, weights, branching
algorithms, etc, and just rely on some sort of Metropolis-style
accept/reject algorithm to produce a correct distribution of walkers.
Within a quantum Hamiltonian framework, a way is known to do this,
namely the Path Integral Monte Carlo (PIMC) approach.\cite{ceperley} We conclude
that the PIMC approach may be better suited than GFMC to the study of
large and complicated lattice Hamiltonian systems. 

\section*{Acknowledgments} This work is supported by the Australian 
Research Council. Calculations were performed on the SGI Power Challenge 
Facility at the New South Wales Centre for Parallel Computing and the 
Fujitsu VPP300 vector machine at the Australian National Universtiy 
Supercomputing Facility: we are grateful for the use of these facilities.

\begin{figure}[htbp]
\caption{
Ground-state energy per site graphed against $1/L^{4}$, where L is the
lattice size: a) at coupling $\lambda = 3.0$, b) at coupling $\lambda =
5.0$. The lines are merely to guide the eye.
}
\label{fig1}
\end{figure}
\begin{figure}[htbp]
\caption{
The bulk ground-state energy per site graphed against coupling
$\lambda$. The points are our Monte Carlo estimates; the solid line
represents earlier ELCE estimates[19]; and the dashed line represents the
asymptotic weak-coupling behaviour.
}
\label{fig2}
\end{figure}
\begin{figure}[htbp]
\caption{
Estimated value for the mean plaquette $W(1,1)$ as a function of the
variational parameter $c$, for $L= 6, \lambda = 5.0$.
}
\label{fig3}
\end{figure}
\begin{figure}[htbp]
\caption{
The mean plaquette $W(1,1)$ for the bulk system graphed against coupling
$\lambda$. The solid line represents the strong-coupling series
expansion[19], and the dashed line the asymptotic weak-coupling
behaviour.
}
\label{fig4}
\end{figure}
\begin{figure}[htbp]
\caption{
The string tension $Ka^{2}$ graphed against coupling $\beta$. The
circles are obtained from ELCE estimates of the axial string 
tension[19];
the triangles are Monte Carlo estimates of $R_{2}$.
}
\label{fig5}
\end{figure}

\end{document}